\documentclass[conference]{IEEEtran}
\IEEEoverridecommandlockouts
\usepackage{cite}
\usepackage{amsmath,amssymb,amsfonts}
\usepackage{algorithmic}
\usepackage{graphicx}
\newtheorem{theorem}{Theorem}
\newtheorem{lemma}{Lemma}
\newtheorem{corollary}{Corollary}[theorem]
\usepackage{textcomp}
\usepackage{xcolor}
\def\BibTeX{{\rm B\kern-.05em{\sc i\kern-.025em b}\kern-.08em
		T\kern-.1667em\lower.7ex\hbox{E}\kern-.125emX}}
\usepackage{subfigure}

\title{sparse array for ISAC system}
\begin{document}
\bibliographystyle{IEEEtran}
\title{Can Sparse Arrays Outperform Collocated Arrays for Future Wireless Communications?\\}
\author{\IEEEauthorblockN{Huizhi Wang\IEEEauthorrefmark{1} and 
		Yong Zeng\IEEEauthorrefmark{1}\IEEEauthorrefmark{2}}
	
	\IEEEauthorblockA{\IEEEauthorrefmark{1}National Mobile Communications Research Laboratory, Southeast University, Nanjing 210096, China}
	\IEEEauthorblockA{\IEEEauthorrefmark{2}Purple Mountain Laboratories, Nanjing 211111, China}
	\IEEEauthorblockA{Email: wanghuizhi@seu.edu.cn, yong\underline{~}zeng@seu.edu.cn.}
}

\maketitle
\begin{abstract}
\textbf{Multiple-input multiple-output (MIMO) has become a key technology for contemporary wireless  communication systems. For typical MIMO systems, antenna arrays are separated by half of the signal wavelength, which are termed {\emph {collocated arrays}}. In this paper, we ask the following question: For future wireless communication systems, is it possible to achieve better performance than collocated arrays by using  sparse arrays, whose element spacing is larger than half wavelength? The answer to this question is not immediately clear since while sparse arrays may achieve narrower beam for the main lobe, they also generate undesired grating lobes. In this paper, we show that the answer to the above question is affirmative. To this end, we first provide an insightful explanation by investigating the key properties of beam patterns of sparse and collocated arrays, together with the typical distribution of spatial angle difference $\Delta$, which all critically impact the inter-user interference (IUI). In particular, we show that sparse arrays are less likely to experience severe IUI than collocated arrays, since the probability of $\Delta$ typically reduces with the increasing of $|\Delta|$. This naturally helps to reject those higher-order grating lobes of sparse arrays, especially when users are densely located. Then we provide a rigorous derivation of the achievable data rate for sparse and collocated arrays, and derive the condition under which sparse arrays strictly outperform collocated counterparts. Finally, numerical results are provided to validate our theoretical studies.}
\end{abstract}

\section{Introduction}
Multiple-input multiple-output (MIMO) has become a key technology in contemporary wireless communication systems, thanks to its fundamental spatial multiplexing gain and diversity gain over single-antenna systems\cite{b19}. Over the past few decades, MIMO technology has been tremendously advanced from small MIMO in the fourth-generation (4G) mobile communication networks to massive MIMO in 5G\cite{b21}. Looking forward towards 6G, different pathways have been pursued for the further evolution of MIMO technology. The first one is to continuously increase the number of antenna elements and physical/electrical size of antenna array beyond current massive MIMO systems, leading to the paradigm of extremely large-scale MIMO (XL-MIMO) communications\cite{b2,b29,b102,b103}. Another pathway is to significantly reduce the spacing of adjacent array elements, say using sub-wavelength structure, instead of the typical half-wavelength spacing for standard discrete antenna arrays. With the advancement of antenna design technology such as metamaterials, such methods lead to communication systems with semi-passive devices such as reconfigurable intelligent surface (RIS) \cite{b101} or large intelligent surface (LIS) \cite{b99}. In the extreme case when inter-element spacing vanishes, the conventional discrete antenna arrays would reduce to continuous surfaces \cite{b23} or holographic MIMO \cite{b99}. Note that while continuously increasing the array size in XL-MIMO leads  to larger array aperture, it increases the hardware cost, power consumption, and signal processing complexity.  On the other hand,  using semi-passive subarray structures needs to address the practical mutual coupling issues, and the array aperture is typically limited unless  a huge number  of passive elements are used. 

An alternative method to enhance the array aperture, yet without having to increase the number of array elements, is to use {\emph {sparse arrays}}, whose element spacing is larger than half of the signal wavelength. Note that for the standard {\emph {collocated arrays}}, the value of half-wavelength is chosen for antenna spacing due to several reasons. On the one hand, an element spacing smaller than half wavelength  may lead to non-negligible mutual coupling, which severely affects radiation pattern and diversity behavior\cite{b91}\cite{b92}. On the other hand, antenna array with interelement-spacing larger than half wavelength is known to suffer from undesired grating lobes, for which equally strong power will be directed towards other directions, besides the desired beamforming direction. For communication systems, two users located within grating lobes of each other will suffer from severe inter-user interference (IUI). For radar sensing systems, grating lobes are also undesired since they cause ambiguity for direction estimation. 

On the other hand, sparse arrays also have their own merits, since the beamwidth of the main lobe is narrower than the collocated counterpart, thanks to the enlarged total aperture. This implies that they are expected to provide better interference mitigation for closely located users, which is appealing for hot-spot areas. Furthermore, sparse transmit or receive arrays have also been used to create virtual MIMO in radar sensing systems, i.e., by using only $M+N$ physical array elements while achieving $MN$ virtual aperture\cite{b27}. Based on the above discussions, it is not immediately clear whether it is possible for sparse arrays to outperform the standard collocated arrays for  future wireless communication systems, which motivates our current work.   

In this paper, we show that the answer to the above question is affirmative. To begin with, the beam patterns of sparse and collocated arrays are analyzed, as well as the distribution of spatial angle difference $\Delta$ between two randomly distributed users, which all play crucial roles in determining the level of IUI. We show that the probability of $\Delta$ typically decreases as $|\Delta|$ increases, which effectively mitigates higher-order grating lobes of sparse arrays. Therefore, sparse arrays are less susceptible to severe IUI compared to collocated arrays, especially for densely located users. Furthermore, we derive the probability distribution of achievable data rate for both sparse and collocated arrays, and the conditions under which sparse arrays will outperform their collocated counterparts. 
\section{SYSTEM MODEL}
\begin{figure}[htbp]
	\setlength{\abovecaptionskip}{0.1cm}
	\setlength{\belowcaptionskip}{-0.1cm}
	\centerline{\includegraphics[width=0.3\textwidth]{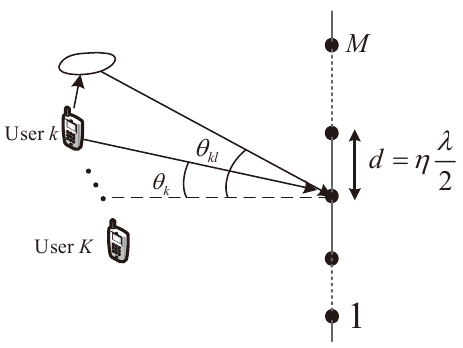}}
	\caption{Wireless communication systems using ULAs with inter-element spacing $d=\eta \lambda/2$, where $\eta=1$ corresponds to standard collocated arrays, and $\eta>1$ for sparse arrays.}
	\label{sparse}
\end{figure}
As shown in Fig. \ref{sparse}, we consider a multi-user wireless communication system, where a base station (BS) equipped with a $M$-element uniform linear array (ULA) serves $K$ single-antenna users. Let $d=\eta\frac{\lambda}{2}$ denote the separation of adjacent array elements, where $\lambda$ is the signal wavelength and $\eta\geq 1$ denotes the antenna separation parameter. Two different array setups will be considered according to the value of $\eta$, namely the standard {\emph {collocated array}} with $\eta=1$, and {\emph {sparse array}} with $\eta>1$. 

Let ${\bf {h}}_k\in \mathbb{C}^{M\times 1}$ denotes the channel vector between user $k$ and the BS, which can be expressed as
\begin{equation}
	\setlength\abovedisplayskip{2pt}
	\setlength\belowdisplayskip{2pt}
	\begin{aligned}
{{\bf{h}}_k} = \sum\limits_{l = 1}^{L_k} \beta_{kl} {\bf a}(\theta_{kl}) ,
	\end{aligned}
	\label{h}
\end{equation}
where $L_k$ denotes the number of multi-paths of user $k$, $\beta_{kl}$ denotes the complex-valued path gain of the $l$th path of user $k$, and ${\bf{a}}({\theta _{kl}}) = {\left[ {1,{e^{j\pi \eta \sin {\theta _{kl}}}},...,{e^{j(M - 1)\pi \eta \sin {\theta _{kl}}}}} \right]^T}$ denotes the array response vector, with $\theta_{kl}$ denoting the angle-of-arrival (AoA) of the $l$-th path of user $k$. For the special case of LoS dominating channel, we have $L_k=1$ and 
\begin{equation}
	\setlength\abovedisplayskip{2pt}
	\setlength\belowdisplayskip{2pt}
	\begin{aligned}
		{{\bf{h}}_k} = \beta_{k} {\bf a}(\theta_{k})=\beta_k{\left[ {1,{e^{j\pi \eta \sin {\theta _{k}}}},...,{e^{j(M - 1)\pi \eta \sin {\theta _{k}}}}} \right]^T},
	\end{aligned}
	\label{hlos}
\end{equation}
where $\beta_k$ and $\ \theta_k  \in \left[ { - {\pi  \mathord{\left/
			{\vphantom {\pi  {2,{\pi  \mathord{\left/
								{\vphantom {\pi  2}} \right.
								\kern-\nulldelimiterspace} 2}}}} \right.
			\kern-\nulldelimiterspace} {2,{\pi  \mathord{\left/
					{\vphantom {\pi  2}} \right.
					\kern-\nulldelimiterspace} 2}}}} \right]\ $are the path gain and direction of user $k$, respectively.

For uplink communication, the received signal at the BS can be expressed as
\begin{equation}
	\setlength\abovedisplayskip{2pt}
	\setlength\belowdisplayskip{2pt}
	\begin{aligned}
{{\bf{y}}} = \sum\limits_{i = 1}^K {{{\bf{h}}_i}} \sqrt {{P_i}} {s_i} + {\bf{n}},
	\end{aligned}
	\label{receive}
\end{equation}
where$\ s_i $ and $P_i$ denote the information-bearing symbol and transmit power of user $i$, respectively, with $E[|s_i|^2]=1$, $\ {\bf n} \sim \mathcal{CN}(\mathbf 0,{\sigma ^2\mathbf{I}_M})\ $denotes the additive white Gaussian noise (AWGN). To detect the signal for user $k$, a receive beamforming vector $\ {{\bf{v}}_k}\ \in \mathbb{C}^{M\times 1}$ with $||{{\bf{v}}_k}||=1$ is used. The resulting signal is
\begin{equation}
	\setlength\abovedisplayskip{2pt}
	\setlength\belowdisplayskip{2pt}
	\begin{aligned}
{y_k} ={\bf{v}}_k^H {{\bf{y}}}={\bf{v}}_k^H{{\bf{h}}_k}\sqrt {{P_k}} {s_k} + {\bf{v}}_k^H\sum\limits_{i = 1,i \ne k}^K {{{\bf{h}}_i}} \sqrt {{P_i}} {s_i} + {\bf{v}}_k^H{\bf{n}}.
	\end{aligned}
	\label{bf}
\end{equation}

The signal-to-interference-plus-noise ratio (SINR) for user $k$ is
\begin{equation}
	\setlength\abovedisplayskip{2pt}
	\setlength\belowdisplayskip{1pt}
	\begin{aligned}
{\gamma _k} = \frac{{{P_k}{{\left| {{\bf{v}}_k^H{{\bf{h}}_k}} \right|}^2}}}{{\sum\limits_{i = 1,i \ne k}^K {{P_i}{{\left| {{\bf{v}}_k^H{{\bf{h}}_i}} \right|}^2}}  + {\sigma ^2}}}, k=1,...,K.
		\label{sinr}
	\end{aligned}
\end{equation}
It is well known that three classical beamforming methods can be used to achieve different balance between complexity and performance, namely, maximal-ratio combining (MRC), zero-forcing (ZF), and minimum mean-square error (MMSE). Their corresponding SINRs can be expressed as\cite{b96}
\begin{equation}
	\setlength\abovedisplayskip{2pt}
	\setlength\belowdisplayskip{1pt}
	\begin{aligned}
	{\gamma _{k,{\rm{MRC}}}} & = \frac{{{{ P}_k}{{\left\| {{{\bf{h}}_k}} \right\|}^2}}}{{\sum\limits_{i = 1,i \ne k}^K {{{ P}_i}{{\left| {\frac{{{\bf{h}}_k^H}}{{\left\| {{{\bf{h}}_k}} \right\|}}{{\bf{h}}_i}} \right|}^2}}  + {\sigma ^2}}},\\
	{\gamma _{k,{\rm{ZF}}}} &= \frac{{P}_k}{\sigma^2}{\bf{h}}_k^H\left( {{{\bf{I}}_M} - {{\overline {\bf{H}} }_k}{{\left( {\overline {\bf{H}} _k^H{{\overline {\bf{H}} }_k}} \right)}^{ - 1}}\overline {\bf{H}} _k^H} \right){{\bf{h}}_k},\\
	{\gamma _{k,{\rm{MMSE}}}} &= \frac{{P}_k}{\sigma^2}{\bf{h}}_k^H{\bf{C}}_k^{ - 1}{{\bf{h}}_k},
		\label{bf3}
	\end{aligned}
\end{equation}
where ${\overline {\bf{H}} _k} \triangleq [{{\bf{h}}_1},...,{{\bf{h}}_{k - 1}},{{\bf{h}}_{k + 1}},...,{{\bf{h}}_K}]$, and ${{\bf{C}}_k} \triangleq \sum\nolimits_{i \ne k} {{{\tilde P}_i}{{\bf{h}}_i}{\bf{h}}_i^H}  + {{\bf{I}}_M}$.
The achievable rate in bits/second/Hz of user $k$ is thus
\begin{equation}
	\setlength\abovedisplayskip{2pt}
	\setlength\belowdisplayskip{1pt}
	\begin{aligned}
		R_k = {{{\log }_2}(1 + {\gamma _k})}.
		\label{sumrate}
	\end{aligned}
\end{equation}
It is observed from (\ref{hlos}) and (\ref{bf3}) that the inter-element spacing parameter $\eta$ critically affects the performance via the array response vector ${\bf a}(\theta_k)$. In this paper, we ask the following question: Is it possible for sparse arrays with $\eta>1$ outperform the standard collocated arrays with $\eta=1$? 

To answer the above question, we first consider the very basic LoS-dominating scenario with MRC beamforming, and explain how sparse and collocated arrays differ by examining their beam patterns, together with the distributions of spatial angle difference of two randomly located users. After that, the achievable data rate of sparse and collocated  arrays are derived. Finally, numerical results are provided for the generic multi-path scenarios for all the three aforementioned beamforming techniques.
\begin{figure*}[htbp]
	\centering
	\subfigure[collocated ULA ($\eta=1$)]{
		\includegraphics[width=5cm]{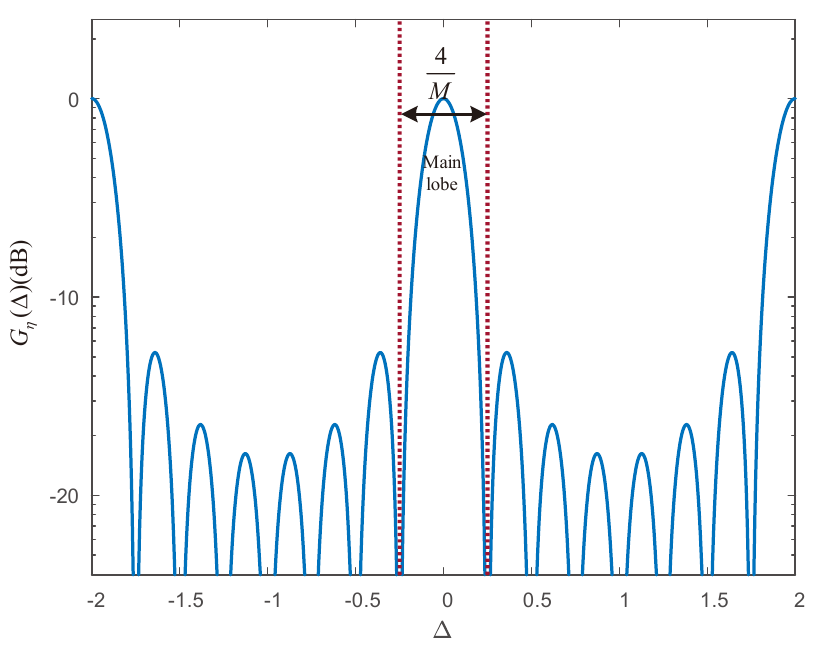}
	}
	\quad
	\subfigure[sparse ULA ($\eta=4$)]{
		\includegraphics[width=5.1cm]{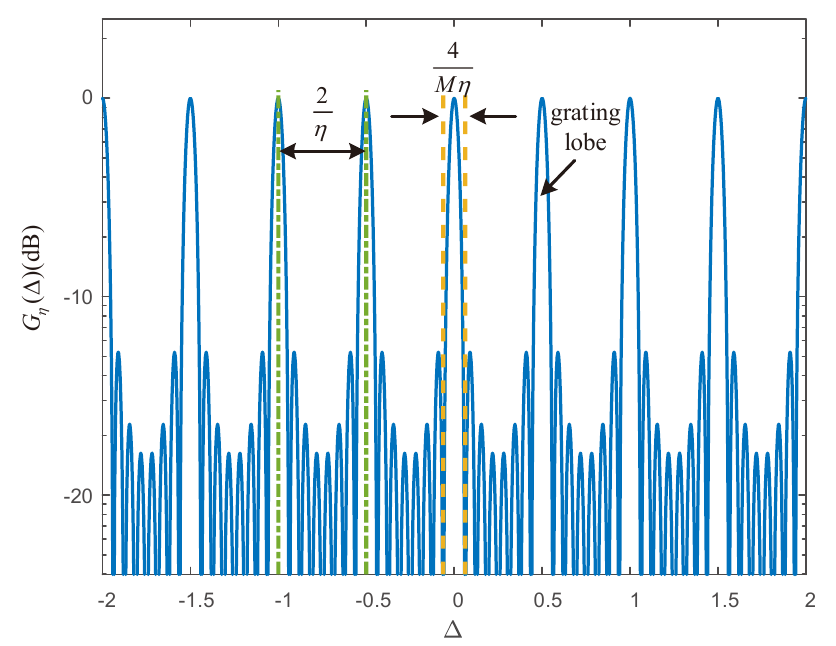}
	}
	\quad
	\subfigure[PDF of spatial angle difference $\Delta$]{
		\includegraphics[width=5.1cm]{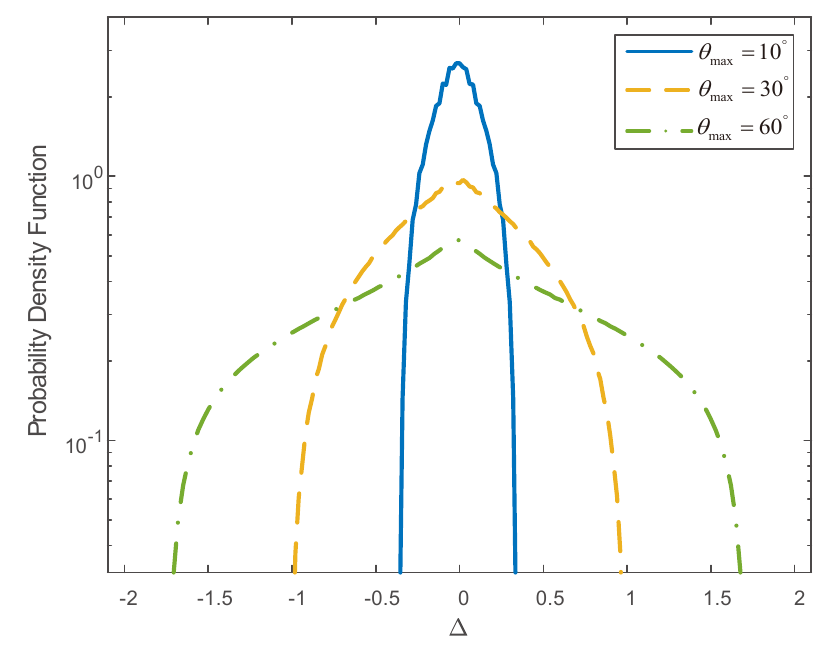}
	}
	\caption{Beam patterns of collocated and sparse ULAs, together with the PDF of spatial angle difference $\Delta$. }
	\label{pattern}
\end{figure*}
\section{How Sparse and Collocated Arrays Differ?}
For the basic LoS-dominating scenario, by substituting (\ref{hlos}) into (\ref{bf3}), the SINR for the low-complexity MRC beamforming is simplified as
\begin{equation}
	\setlength\abovedisplayskip{2pt}
	\setlength\belowdisplayskip{1pt}
	\small
	\begin{aligned}
{\gamma _k} = \frac{{{{ P}_k}{{\left| {{\beta _k}} \right|}^2}M}}{{{{\left| {{\beta _i}} \right|}^2}M\sum\limits_{i = 1,i \ne k}^K {{{ P}_i}{{\left| {\frac{{{\bf{h}}_k^H{{\bf{h}}_i}}}{{\left\| {{{\bf{h}}_i}} \right\|\left\| {{{\bf{h}}_k}} \right\|}}} \right|}^2}}  + \sigma^2}} = \frac{{{{\bar P}_k}M}}{{M\sum\limits_{i = 1,i \ne k}^K {{{\bar P}_i}} {\rho _{ki}} + 1}},
		\label{SINR}
	\end{aligned}
\end{equation}
where ${{\bar P}_k} \triangleq \frac{{{{\left| {{\beta _k}} \right|}^2}{P_k}}}{{{\sigma ^2}}}$, and ${\rho _{ki}} \buildrel \Delta \over = \frac{{{{\left| {{\bf{h}}_k^H{{\bf{h}}_i}} \right|}^2}}}{{{{\left\| {{{\bf{h}}_k}} \right\|}^2}{{\left\| {{{\bf{h}}_i}} \right\|}^2}}}$ denotes the absolute square of the channel correlation between user $k$ and user $i$. For LoS-dominating channels, it follows from (\ref{hlos}) that
\begin{equation}
	\setlength\abovedisplayskip{2pt}
	\setlength\belowdisplayskip{1pt}
	\small
	\hspace{-1ex}
	\begin{aligned}
			\rho_{ki}={\left| {\frac{1}{M}\sum\limits_{m = 0}^{M - 1} {{a}_m^*({\theta _i}){{a}_m}({\theta _k})} } \right|^2} = {\left| {\frac{{\sin \left[ {\frac{\pi }{2}M\eta {\Delta _{ik}}} \right]}}{{M\sin \left[ {\frac{\pi }{2}\eta {\Delta _{ik}}} \right]}}} \right|^2}\triangleq G_\eta(\Delta_{ik}),
			\label{rho}
	\end{aligned}
\end{equation}
where ${\Delta _{ik}} = \sin {\theta _k} - \sin {\theta _i}\in [-2, 2]$ is the spatial angle difference. It follows from (\ref{SINR}) and (\ref{rho}) that the SINR of each user $k$ critically depends on the beam pattern $G_\eta(\Delta_{ik})$, which is a function of the spatial angle difference $\Delta_{ik}$ with parameter $\eta$. Following similar analysis as \cite{b104}, we have the following properties for the function $G_\eta(\Delta)$: 

\emph{Beamwidth of main lobe:} The null points of the main lobe for $G_\eta(\Delta)$ can be obtained by letting $\frac{\pi }{2 }M\eta{\Delta } =  \pm \pi$. Therefore, the beamwidth of the main lobe is
\begin{equation}
	\setlength\abovedisplayskip{2pt}
	\setlength\belowdisplayskip{1pt}
	\begin{aligned}
BW =  \frac{4}{{M\eta}},
\label{nnb}
	\end{aligned}
\end{equation}
which decreases with $\eta$. This implies that two users that have small spatial angle difference $\Delta_{ik}$, may suffer from severe IUI for collocated arrays, but weak interference for sparse ULA. From this perspective, sparse arrays seem to be preferred.  

\emph{ Grating lobes:} It is well known that when the inter-element spacing is larger than half wavelength, undesired grating lobes that have equal amplitude and width as the main lobe will be generated. The spatial directions of grating lobes can be obtained by letting $\frac{\pi }{2 }\eta{\Delta } = n\pi ,n =  \pm 1, \pm 2..., \pm \left\lfloor \eta  \right\rfloor $, and thus the $n$-th grating lobe will appear at
\begin{equation}
	\setlength\abovedisplayskip{2pt}
	\setlength\belowdisplayskip{1pt}
	\begin{aligned}
{\Delta } = \frac{{2n}}{\eta}, n =  \pm 1, \pm 2..., \pm \left\lfloor \eta  \right\rfloor.
	\end{aligned}
\end{equation}
For $M=8$, Fig.\ref{pattern} (a) and (b) plot the beam pattern function $G_\eta(\Delta)$ for $\eta=1$ and  $\eta=4$, respectively. It is observed that on the one hand, a sparse ULA is able to reduce IUI for users with small angular separations. On the other hand, it may result in stronger IUI than collocated arrays when two users lie in the grating lobe of each other. Therefore, it is not immediately clear whether there is any chance for sparse arrays to outperform the standard collocated arrays.  

In order to resolve the above issue, in Fig.\ref{pattern} (c), we numerically plot the distribution of $\Delta$ when the AoAs of the users are uniformly distributed as $\theta_k\in [-\theta_{\max},\theta_{\max}], \forall k$. It is observed that for all the three values of $\theta_{\max}$ considered, the probability of $\Delta$ decreases with the increase of $|\Delta|$. This is a quite appealing feature for sparse arrays, since the non-uniform distribution of $\Delta$, even with uniform user angle distribution, may act like a natural filter, to automatically reject those higher-order grating  lobes of sparse arrays. As a result, compared to collocated arrays, it becomes less likely that $\Delta$ will fall into the severe IUI region, i.e.,  either in the main lobe or grating lobe.  As a concrete example, for $\theta_{\max}=10^\circ,$ the spatial angle difference $\Delta$ is concentrated within the range $\left| \Delta  \right| < 0.36$. Within this range, only the main lobe will cause severe IUI for both collocated and sparse arrays. Furthermore, since sparse arrays have much narrower main lobe than the collocated counterpart, they are less likely to suffer from severe IUI and hence a better performance is expected.

In summary, the above discussions demonstrate that thanks to the typically non-uniform distribution of the spatial angle difference $\Delta$, sparse arrays may outperform the standard collocated arrays due to the automatic suppression of high-order grating lobes, especially when users are densely located, i.e., with small $\theta_{\max}$.  In order to make such intuitive understandings concrete, in the following, we provide rigorous theoretical analysis for the probability distribution of the achievable user data rate (\ref{sumrate}) for randomly located users, and provide the performance comparison for sparse and collocated arrays. 
\section{Distribution of Achievable Rate for Sparse and Collocated Arrays}
Note that (\ref{rho}) is difficult for mathematical analysis. Therefore, similar to the two-lobe approximation\cite{b97}, we use the following approximation:
\begin{equation}
	\setlength\abovedisplayskip{2pt}
	\setlength\belowdisplayskip{1pt}
	\hspace{-1ex}
	\begin{aligned}
		G_\eta(\Delta) \approx \begin{cases}
			{G_m},&- t \le {\Delta} - \frac{{2n}}{\eta } \le t,n = 0,...,\pm\left\lfloor \eta  \right\rfloor \\
			{G_s},&{\rm{otherwise}}
		\end{cases},
	\label{twolobe}
	\end{aligned}
\end{equation}
where $2t=\frac{2\alpha}{M\eta}$ denotes the effective beamwidth of the main lobe or grating lobe, with $\alpha\in[0,2]$. $G_m$ and $G_s$ are the gain of main/grating lobe and side lobe, respectively, with $G_s<G_m$. The value of $\alpha$, $G_m$ and $G_s$ can be obtained by performing a least-square based curve fitting. 

For simplicity, we assume that ${{\bar P}_k}={{\bar P}},\forall k.$ Furthermore, $X_k\triangleq\sin\theta_k$ follows i.i.d distribution with the support $[-\sin\theta_{\max},\sin\theta_{\max}]$, and its PDF is denoted as $f_X$. Therefore, the joint PDF of $X_i$ and $X_k$ is
\begin{equation}
	\setlength\abovedisplayskip{2pt}
	\setlength\belowdisplayskip{1pt}
	\begin{aligned}
f\left( {{X_i},{X_k}} \right) = \begin{cases}
	f_{X_i}f_{X_k}, &- \sin\theta_{\max} \le {X_i},{X_k} \le \sin\theta_{\max}\\
0,&{\rm{otherwise}}
\end{cases}.
	\end{aligned}
\end{equation}
Furthermore, the probability that two users lie in the main lobe or grating lobe, denoted as $p$, can be expressed as
\begin{equation}
	\setlength\abovedisplayskip{2pt}
	\setlength\belowdisplayskip{1pt}
	\small
\begin{aligned}
	p \triangleq \mathrm{Pr}(G_\eta(\Delta_{ik})  = {G_m}) \approx \iint_{\mathbb{D}}{f_{X_i}f_{X_k}}d{X_i}d{X_k},
	\label{p1}
	\end{aligned}
\end{equation}
where ${\mathbb{D}} = \{ ({X_i},{X_k})| - \sin\theta_{\max} \le {X_i},{X_k} \le \sin\theta_{\max}, - t \le {X_i} - {X_k}-\frac{2n}{\eta}\le t, n = 0,...,\pm\left\lfloor \eta  \right\rfloor\}$. 

\begin{theorem}\label{t1}
With the approximation in (\ref{twolobe}), the cumulative distribution function (CDF) of the achievable data rate in (\ref{sumrate}) can be expressed as
	\begin{equation}
		\setlength\abovedisplayskip{2pt}
		\setlength\belowdisplayskip{1pt}
		\hspace{-1ex}
		\begin{aligned}
			\begin{array}{l}
				F = \mathrm{Pr}(R\leq \bar R)= 1 - \sum\limits_{q = 0}^N {C_{K - 1}^q}{{{{p}} }^q}{{(1 - {p})}^{(K - 1) - q}},
			\end{array}
		\end{aligned}
	\label{bi}
	\end{equation}
where $p$ is defined in (\ref{p1}), $C_{K - 1}^q = \frac{{(K - 1)!}}{{q!(K - 1 - q)!}}$ is the binomial coefficient, $N = \left\lfloor {\frac{Y-(K-1)G_s}{{G_m - G_s}}} \right\rfloor,$ and $Y = \frac{1}{{{2^{\bar R}} - 1}} - \frac{1}{{\bar PM}}$.
\end{theorem}
\begin{IEEEproof}
	Please refer to Appendix A.
\end{IEEEproof}

\begin{lemma}
When $K\to\infty$, the CDF of the achievable data rate in (\ref{sumrate}) can be expressed as
	\begin{equation}
		\setlength\abovedisplayskip{2pt}
		\setlength\belowdisplayskip{1pt}
		\begin{aligned}
		F = 1 - \frac{1}{2}\left[ {1 + erf\left( {\frac{{Y - \mu }}{{\sqrt {2\nu } }}} \right)} \right],
		\end{aligned}
	\label{no}
	\end{equation}
	where $erf(\cdot)$ is the error function, $\mu  =(K-1) ({G_s} + ({G_m} - {G_s})p) $, and $\nu  = (K-1){({G_m} - {G_s})^2}p(1 - p)$.
	\label{l1}
\end{lemma}
\begin{IEEEproof}
When $K$ is large, the sum of the IUI coefficients approaches to a Gaussian distribution, i.e., $\sum\limits_{i = 1,i \ne k}^K {{\rho _{ki}}} \sim {\mathcal {N}}(\mu ,\nu ),$ with $ \mu  = \left( {K - 1} \right)\left( {{G_s} + ({G_m} - {G_s})p} \right),\nu  = \left( {K - 1} \right){({G_m} - {G_s})^2}p(1 - p)$ denoting the expectation and covariance of $ \sum\limits_{k = 1,k \ne i}^K {{\rho _{ik}}} $. Then the CDF in (16) can be readily obtained.
\end{IEEEproof}
The expressions in (\ref{bi}) and (\ref{no}) show that the probability $p$ defined in (\ref{p1}) critically affects the rate CDF. In the following, by assuming $X_i \sim U( - \sin {\theta _{\max }},\sin {\theta _{\max }}),\forall i,$ the closed-form expression of $p$ is derived. 
\begin{theorem}
For $M$-element ULA with element spacing parameter $\eta$, the probability $p$ in (\ref{p1}) that two users lie in the main/grating lobes is
	\begin{equation}
		\setlength\abovedisplayskip{2pt}
		\setlength\belowdisplayskip{1pt}
		\small
		\begin{aligned}
			&{p(\eta)} \approx \\
			&\begin{cases}
				{\frac{{\alpha \left( {\left( {2{n_{\max }} + 1} \right)\sin {\theta _{\max }} - \frac{\alpha }{{4M\eta }} - \frac{{{n_{\max }}\left( {{n_{\max }} + 1} \right)}}{\eta }} \right)}}{{{{\sin }^2}{\theta _{\max }}M\eta }}},&\sin {\theta _{\max}}\ge {{\frac{\alpha }{{2M\eta }}}} \\
				1,&\sin {\theta _{\max}}<{{\frac{\alpha }{{2M\eta }}}} 
			\end{cases}.
		\end{aligned}
		\label{ps}
	\end{equation}
	where ${n_{\max}} = \left\lfloor {\eta\sin \theta_{\max}  - \frac{{\alpha}}{2M}} \right\rfloor. $
	\begin{IEEEproof}
		Please refer to Appendix B.
	\end{IEEEproof}
	\label{sp}
\end{theorem}
Theorem \ref{sp} shows that $p$ decreases with fluctuation as $\theta_{\max}$ increases. To gain more insights, we consider the special case of collocated arrays by letting $\eta=1$ in Theorem \ref{sp}:
\begin{corollary}
For collocated ULAs with $\eta=1$, $p$ in (\ref{ps}) reduces to 
	\begin{equation}
		\setlength\abovedisplayskip{2pt}
		\setlength\belowdisplayskip{1pt}
		\small
		\begin{aligned}
			{p_\mathrm{col}} &\approx \begin{cases}
				\frac{{4\alpha M\sin {\theta _{\max }} - {\alpha ^2}}}{{4{{\sin }^2}{\theta _{\max }}{M^2}}},&\sin {\theta _{\max}}\ge {{\frac{\alpha }{{2M }}}}\\
				1,&\sin {\theta _{\max}}<{{\frac{\alpha }{2M}}}
			\end{cases}.
		\end{aligned}
		\label{pc}
	\end{equation}
\label{c1}
\end{corollary}
Furthermore, we compare $p(\eta) $ with $\eta>1$ and ${p_\mathrm{col}}$ to show the performance difference between sparse and collocated arrays. 
\begin{theorem}
For $\eta>1$, we have $p(\eta)=p_\mathrm{col}$ when $\sin {\theta _{\max }} \le \sin\underline{\theta}$, $p(\eta)<p_\mathrm{col}$ when $\sin {\theta _{\max }} \in \left( \sin\underline{\theta},\sin\overline{\theta}\right)$, and $p(\eta)>p_\mathrm{col}$ when $\sin {\theta _{\max }} \ge \sin\overline{\theta}$, where $\sin\underline{\theta}\triangleq \frac{\alpha }{{2M\eta }}$ and $\sin\overline{\theta}\triangleq	\frac{{\eta  + \sqrt {{\eta ^2} - \frac{\alpha }{M}({\eta ^2} + 1 - \frac{\alpha }{M})} }}{{2\eta }}. $
	\label{appendix b}
\end{theorem}
\begin{IEEEproof}
Please refer to Appendix C.
\end{IEEEproof}
Theorem \ref{appendix b} implies that when $K$ is large so that $\sum\limits_{i = 1,i \ne k}^K {{\rho _{ki}}} $ approach to $(K - 1)\left( {\left( {{G_m} - {G_s}} \right)p + {G_s}} \right)$ due to the Law of Large Numbers, sparse arrays strictly outperform collocated arrays when $\sin {\theta _{\max }} \in \left( \sin\underline{\theta},\sin\overline{\theta}\right).$ For example, when $M=16,\eta=5.5$, and $\alpha=1.6$, we have $\underline{\theta}\approx 0.5^\circ$ and $\overline{\theta}\approx77^\circ$, which is a wide region for practical communication systems.
\section{Simulation results}
Unless otherwise stated, we set $M=32,\eta=4$, and $\bar P = 20$dB. For the multi-path channel in (\ref{h}), we use the ``one-ring" model, with $L_k=10$ multi-paths, $R=5$m denoting the radius of each ring, and $r=40$m denoting the range of the center of the ring\cite{b88}. The Rician factor is set to be $K_c=20$dB. For the model in (\ref{twolobe}), the values of $\alpha$, $G_m$ and $G_s$ are obtained via curve fitting, given by $\alpha=1.6$, ${G_m} = {\Big| {\frac{{\sin \left( {0.4\pi } \right)}}{{M\sin \left( {0.4\frac{\pi }{M}} \right)}}} \Big|^2},$ and $G_s = 5\times10^{-3}$.

\begin{figure}[htbp]
	\setlength{\abovecaptionskip}{-0.2cm}
	\setlength{\belowcaptionskip}{-0.1cm}
	\centerline{\includegraphics[width=0.45\textwidth]{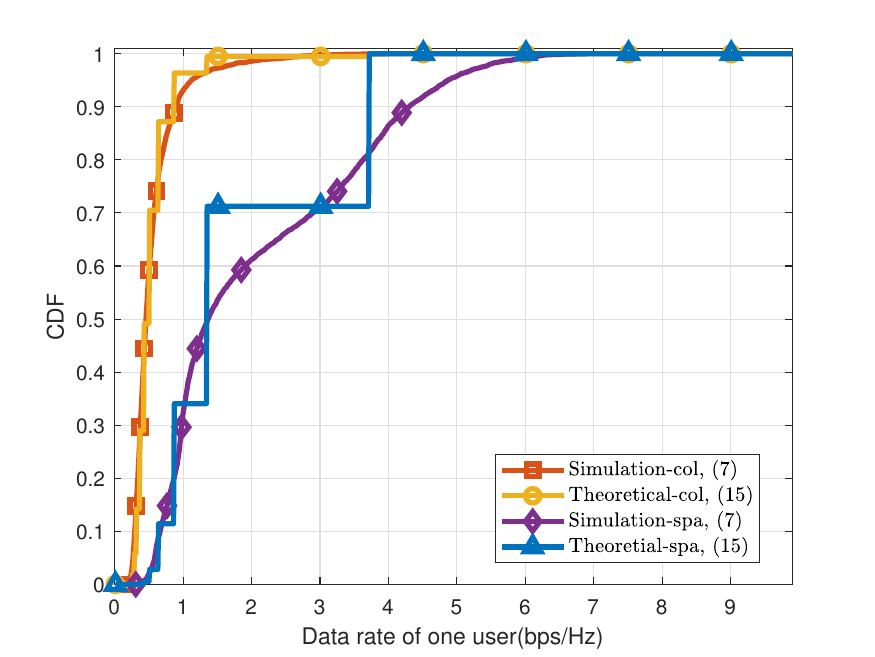}}
	\caption{CDF of data rate for $K=18$ and $\theta_{\max}=10^\circ$.}
	\label{cdf1}
	\vspace{-2ex}
\end{figure}

Fig.\ref{cdf1} shows the simulation and theoretical results for the CDF of data rate in (\ref{bi}) with $K=18$. It can be seen that the developed theoretical results well approximate the simulation results, and the resulting gap may come from the two-lobe beam approximation in (\ref{twolobe}). Besides, considering $90$-percentile data rate, both theoretical and simulation results show that sparse ULA can achieve more than four-times data rate than collocated ULA. 
\begin{figure}[htbp]
	\setlength{\abovecaptionskip}{-0.2cm}
	\setlength{\belowcaptionskip}{-0.1cm}
	\centerline{\includegraphics[width=0.45\textwidth]{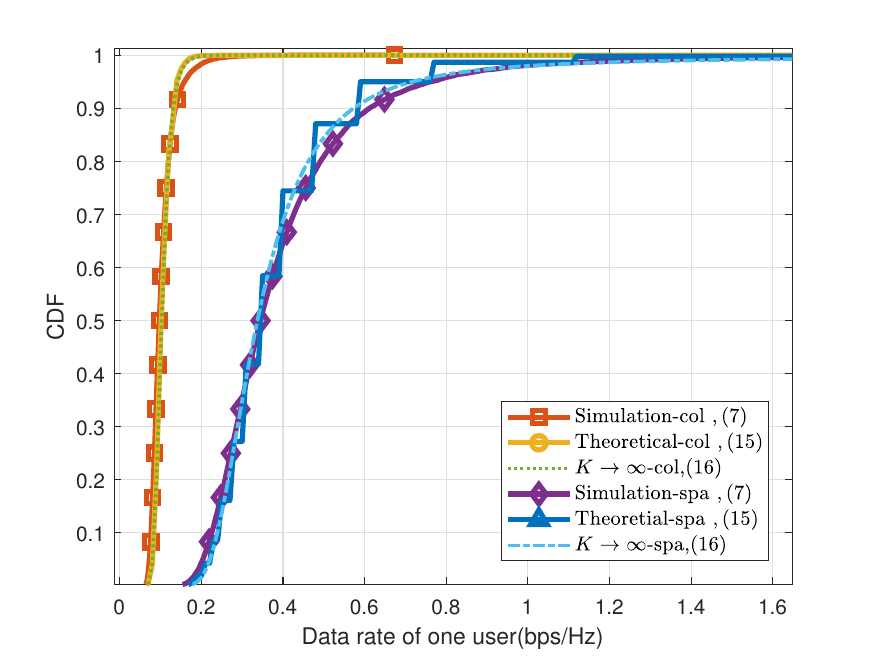}}
	\caption{CDF of data rate for $K=88$ and $\theta_{\max}=10^\circ$.}
	\label{lcdf1}
		\vspace{-2ex}
\end{figure}
\begin{figure}[htbp]
	\setlength{\abovecaptionskip}{-0.2cm}
	\setlength{\belowcaptionskip}{-0.1cm}
	\centerline{\includegraphics[width=0.45\textwidth]{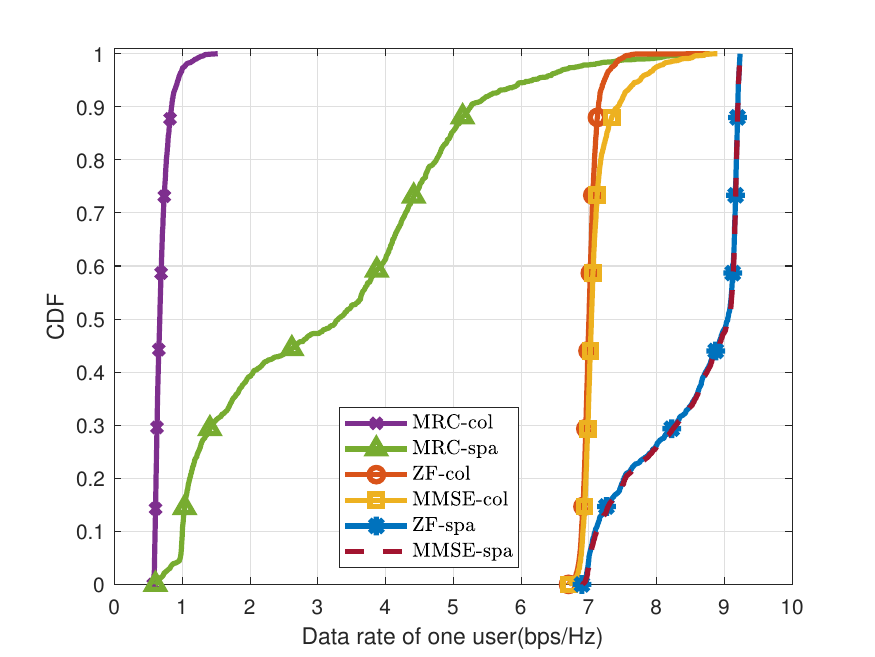}}
	\caption{CDF of data rate for $K=3$ and $\theta_{\max}=6^\circ$.}
	\label{zm}
\end{figure}
Fig.\ref{lcdf1} shows the simulation and theoretical value, together with the Normal approximation in (\ref{no}) when there are $K=88$ users. Compared with Fig. \ref{cdf1}, it can be seen that the achievable data rate decreases with the increase of $K$ since there are more interfering users. It is observed that the sparse ULA still outperforms collocated ULA significantly. Besides, the closed-form expression in Lemma \ref{l1} matches the simulation results quite well. 

Fig.\ref{zm} shows the rate distribution by MRC, ZF, and MMSE beamforming methods, with $M=6$, and $K=3$. For sparse ULA, $\eta=8$ is used. It can be seen that for both collocated and sparse arrays, compared with MRC beamforming, ZF and MMSE can achieve better data rate performance. Besides, for all the three classical beamforming schemes, the sparse ULA significantly outperform the collocated counterpart, thanks to its better IUI mitigation capability for densely located users.

\section{CONCLUSION}
In this paper, we showed that sparse arrays with element spacing greater than half signal wavelength may achieve better performance than the conventional collocated arrays for future wireless communications. By investigating the beam patterns of sparse and collocated arrays, together with the probability distribution of spatial angle difference, we revealed that sparse arrays are less likely to suffer from severe IUI than collocated arrays because of the non-uniform distribution of spatial angle difference. We derived the distribution of achievable data rate for both array types and numerical results were provided to corroborate our theoretical studies.
\section*{APPENDIX A: Proof of Theorem 1}
The CDF of data rate can be obtained based on (\ref{sumrate}) and (\ref{SINR}), with $\bar P_k=\bar P, \forall k.$
\begin{equation}
	\setlength\abovedisplayskip{2pt}
	\setlength\belowdisplayskip{1pt}
	\small
	\begin{aligned}
		F &= \mathrm{Pr}(R_k \le \bar R)= \mathrm{Pr}({\gamma _k} \le {2^{\bar{R}}} - 1)\\
		&= \mathrm{Pr}\Big( {\frac{{{{\bar P}}M}}{{M\sum\limits_{i = 1,i \ne k}^K {{{\bar P}}} {\rho _{ki}} + 1}} \le {2^{\bar{R}}} - 1}\Big )\\
		&=\mathrm{Pr}\Big( {\sum\limits_{i = 1,i \ne k}^K {{\rho _{ki}}}  \ge \frac{1}{{{2^{\bar{R}}} - 1}} - \frac{1}{{\bar PM}}} \Big)\\
		&= 1 - {F_\rho }\Big({\frac{1}{{{2^{\bar{R}}} - 1}} - \frac{1}{{\bar PM}}} \Big),
	\end{aligned}
\end{equation}
where $F_{\rho}$ is the CDF of $\sum\limits_{i = 1,i \ne k}^K {{\rho _{ki}}}$. Note that ${\rho _{ki}}$ is a random variable with binary values $G_m$ and $G_s$, which can be expressed as ${\rho _{ki}} = ({G_m} - {G_s}){X_{ki}} + {G_s},$ where $X_{ki}$ follows Bernoulli distribution. Therefore, the CDF of $\sum\limits_{i = 1,i \ne k}^K {{\rho _{ki}}}$ is
\begin{equation}
	\setlength\abovedisplayskip{2pt}
	\setlength\belowdisplayskip{1pt}
	\begin{aligned}
{F_\rho }(Y) &= {F_\rho }\Big(\sum\limits_{i = 1,i \ne k}^K {{X_{ki}} < } \frac{{Y - (K - 1){G_s}}}{{{G_m} - {G_s}}}\Big) \\
&= \sum\limits_{q = 0}^N {C_{K - 1}^q} {p^q}{(1 - p)^{(K - 1) - q}},
	\end{aligned}
\end{equation}
where $N\leq K-1$ denotes the number trials out of the $K-1$ total experiments so that $\rho$ chooses the value $G_m$, which can be expressed as $N = \left\lfloor {\frac{{Y - (K - 1){G_s}}}{{{G_m} - {G_s}}}} \right\rfloor. $The result in (\ref{bi}) can then be readily obtained.
\vspace{-1.5ex}
\section*{APPENDIX B: Proof of Theorem 2}
When $X_i \sim U( - \sin {\theta _{\max }},\sin {\theta _{\max }}),$ the PDF of $X_i$ is
\begin{equation}
	\setlength\abovedisplayskip{2pt}
	\setlength\belowdisplayskip{1pt}
	\begin{aligned}
		{f_{X_i}} = \begin{cases}
			\frac{1}{{2\sin {\theta_{\max}}}},&- \sin {\theta_{\max}} \le X_i \le \sin {\theta_{\max}}\\
			0,&{\rm{otherwise}}
		\end{cases}.
	\end{aligned}
\end{equation} 
Therefore, the calculation of $p$ in (\ref{p1}) can be reduced to
\begin{equation}
	\setlength\abovedisplayskip{2pt}
	\setlength\belowdisplayskip{1pt}
	\small
	\begin{aligned}
		p = \iint_{\mathbb{D}}{f_{X_i}f_{X_k}}d{X_i}d{X_k}=\frac{S_{\mathbb{D}}}{4\sin^2\theta_{\max}},
	\end{aligned}
\end{equation}
where $S_{\mathbb{D}}$ denotes the area of ${\mathbb{D}} = \{ ({X_i},{X_k})| - \sin\theta_{\max} \le {X_i},{X_k} \le \sin\theta_{\max}, - t \le {X_i} - {X_k}-\frac{2n}{\eta}\le t,n =  \pm 1, \pm 2..., \pm \left\lfloor \eta  \right\rfloor \}.$  

When $n=0$, $p$ can be obtained by calculating the area as
\begin{equation}
	\setlength\abovedisplayskip{2pt}
	\setlength\belowdisplayskip{1pt}
	\small
	\begin{aligned}
		p_0 = \begin{cases}
			1 - \frac{{{{\left( {2\sin {\theta _{\max }} - {{\left( {\frac{\alpha }{M}} \right)}^2}} \right)}^2}}}{{4{{\sin }^2}{\theta _{\max }}}} = \frac{{4t\sin {\theta _{\max }} - {t^2}}}{{4{{\sin }^2}{\theta _{\max }}}},& 2\sin \theta_{\max}\ge t \\
			1,& 2\sin \theta_{\max}<t
		\end{cases}.
	\end{aligned}
	\label{p0}
\end{equation}
When $n> 0$, we need to consider three cases for different $n$.
\emph{case 1:} In this case, we have $ \sin {\theta _{\max }} - t - \frac{{2n}}{\eta } > - \sin {\theta _{\max }}$, i.e., $0 < n < \eta \sin {\theta _{\max }} - \frac{{t\eta }}{2}$. Note that in this case, $t < 2\sin \theta_{\max}, $ and $p_n$ can be obtained as:
	\begin{equation}
		\setlength\abovedisplayskip{2pt}
		\setlength\belowdisplayskip{1pt}
		\small
		\begin{aligned}
	{p_ n } &= \frac{{\frac{1}{2}{{\left( {2\sin {\theta _{\max }} + t - \frac{{2n}}{\eta }} \right)}^2} - \frac{1}{2}{{\left( {2\sin {\theta _{\max }} - t - \frac{{2n}}{\eta }} \right)}^2}}}{{4\sin^2 \theta _{\max}}}\\
	&= \frac{{t\left( {\sin {\theta _{\max }} - \frac{n}{\eta }} \right)}}{{\sin^2 \theta _{\max}}}
		\end{aligned}
	\end{equation}
\emph{case 2:} Similar to case 1, when $ - t - \frac{{2n}}{\eta} <  - 2\sin \theta_{\max} $ and $ t - \frac{{2n}}{\eta} >  - 2\sin \theta_{\max}$, which means $\eta\sin \theta_{\max}   - {t\eta}/{2} < n < \eta\sin \theta_{\max}  + {t\eta}/{2}$, we have ${p_ n } = {{{( {2\sin {\theta _{\max }} + t - \frac{{2n}}{\eta }} )}^2}}/{(8\sin^2 \theta _{\max})}.$

\emph{case 3:} When $n > \eta\sin \theta_{\max}  + \frac{t\eta}{2}$, we have $p_n = 0.$

When $n<0$, we can obtain similar results due to the symmetric property, and the general expression of $p$ can be expressed as
\begin{equation}
	\setlength\abovedisplayskip{2pt}
	\setlength\belowdisplayskip{1pt}
	\small
	\begin{aligned}
{p_ n } = \begin{cases}
	{\frac{{t\left( {\sin {\theta _{\max }} - \frac{{\left| n \right|}}{\eta }} \right)}}{{{{\sin }^2}{\theta _{\max }}}},}&{0 < \left| n \right| < \eta \sin {\theta _{\max }} - \frac{{t\eta }}{2}}\\
{\frac{{{{\left( {\sin {\theta _{\max }} + \frac{t}{2} - \frac{{\left| n \right|}}{\eta }} \right)}^2}}}{{2{{\sin }^2}{\theta _{\max }}}},}&{ - \frac{{t\eta }}{2} < \left| n \right| - \eta \sin {\theta _{\max }} < \frac{{t\eta }}{2}}\\
{0,}&{\left| n \right| > \eta \sin {\theta _{\max }} + \frac{{t\eta }}{2}}
\end{cases}.
	\end{aligned}
\end{equation}
Therefore, considering all the grating lobes and the main lobe, $p$ can be calculated as
\begin{equation}
	\setlength\abovedisplayskip{2pt}
	\setlength\belowdisplayskip{1pt}
	\small
	\begin{aligned}
		&p = {p_0} + \sum {{p_n}}  \\
		&\mathop  \approx \limits^{(a)} \begin{cases}
				\frac{{4t\sin \theta_{\max}  - {t^2} +8t{n_{\max }}\sin {\theta _{\max }} - \sum\limits_{n =  - {n_{\max }}}^{{n_{\max }}} {\frac{{4tn}}{\eta }} }}{{4\sin {\theta_{\max} ^2}}},&\sin \theta_{\max}  > \frac{t}{2}\\
			1,&\sin \theta_{\max}  < \frac{t}{2}
		\end{cases}\\
	& = \begin{cases}
			\frac{{{t}\left( {2{n_{\max}} + 1} \right)\sin {\theta _{\max}} - \frac{{t^2}}{4} - \frac{{{t}{n_{\max}}\left( {{n_{\max}} + 1} \right)}}{\eta }}}{{\sin \theta _{\max}^2}},&\sin \theta_{\max}  > \frac{t}{2}\\
		1,&\sin \theta_{\max}  < \frac{t}{2}
	\end{cases},
	\end{aligned}
\label{p}
\end{equation}
where ${n_{\max}} = \left\lfloor {\eta\sin \theta_{\max}  - \frac{{t\eta}}{2}} \right\rfloor, $ and $(a)$ holds because typically there will be no $n$ that satisfy $\eta\sin \theta_{\max}   - \frac{t\eta}{2} < n < \eta\sin \theta_{\max}  + \frac{t\eta}{2}$ unless $\eta\sin_{\max}$ is an integer, which is difficult to meet in practice. Therefore, case 2 is not considered, and (\ref{ps}) can be obtained by substituting $t=\frac{\alpha}{\eta M}$ into (\ref{p}).
	\vspace{-1.5ex}
\section*{APPENDIX C: Proof of Theorem 3}
In order to compare $p(\eta)$ and $p_\mathrm{col}$ when $\eta>1$, we first define $\Xi (\sin{\theta _{\max }})\triangleq{p_\mathrm{col}} - {p(\eta)}$. Besides, (\ref{ps}) and (\ref{pc}) show that $p(\eta)$ and $p_\mathrm{col}$ are piecewise function with different turning points $\frac{\alpha }{{2M}}$ and $\frac{\alpha }{{2M\eta }}$, when $\eta\ne 1$. Therefore, $\Xi (\sin{\theta _{\max }})$ is also a piecewise function.

\emph{case 1:} When $0<\sin\theta_{\max}\le\frac{\alpha }{{2M\eta }}$, we have
\begin{equation}
	\setlength\abovedisplayskip{2pt}
	\setlength\belowdisplayskip{1pt}
	\begin{aligned}
		\begin{array}{l}
			\Xi (\sin{\theta _{\max }}) = 1 - 1=0.
		\end{array}
	\end{aligned}
	\label{pc2-ps2}
\end{equation}
In this case, $p(\eta)=p_\mathrm{col}$. 

\emph{case 2:} When $\frac{\alpha }{{2M\eta }} < \sin \theta_{\max}  < \frac{\alpha }{{2M }}$, we have $p_\mathrm{col}=1$ and 
\begin{equation}
	\setlength\abovedisplayskip{2pt}
	\setlength\belowdisplayskip{1pt}
	\small
	\begin{aligned}
		&\Xi (\sin{\theta _{\max }})\\
		&= 1 - 	{\frac{{\alpha \left( {\left( {2{n_{\max }} + 1} \right)\sin {\theta _{\max }} - \frac{\alpha }{{4M\eta }} - \frac{{{n_{\max }}\left( {{n_{\max }} + 1} \right)}}{\eta }} \right)}}{{{{\sin }^2}{\theta _{\max }}M\eta }}}\\
		&\mathop  \approx \limits^{(b)}  \frac{{4{{\sin }^2}{\theta _{\max }}{\eta ^2} - {{\left( {\frac{\alpha }{M}} \right)}^2} - 4\frac{\alpha }{M}{\eta ^2}{{\sin }^2}{\theta _{\max }} + {{\left( {\frac{\alpha }{M}} \right)}^3}}}{{4{{\sin }^2}{\theta _{\max }}{\eta ^2}}}\\
		& = \frac{1}{{4{\eta ^2}}}\left[ {{{\left( {\frac{\alpha }{M}} \right)}^2}\left( {\frac{\alpha }{M} - 1} \right){{\left( {\frac{1}{{\sin {\theta _{\max }}}}} \right)}^2} + 4{\eta ^2}\left( {1 - \frac{\alpha }{M}} \right)} \right],		
	\end{aligned}
	\label{pc2-ps1}
\end{equation}
where $(b)$ holds because $n_{\max}=\left\lfloor {\eta \sin {\theta _{\max }} - \frac{\alpha }{{2M}}} \right\rfloor \approx {\eta \sin {\theta _{\max }} - \frac{\alpha }{{2M}}}$. Note that (\ref{pc2-ps1}) is a concave quadratic function, which will decreases with the increase of $\frac{1}{\sin\theta_{\max}}$ and $\min \Xi (\sin{\theta _{\max }})  = \Xi (\frac{\alpha }{{2M\eta }})= 0$. Therefore, we always have $p(\eta)<p_\mathrm{col}$ in this case.

\emph{case 3:} when $\frac{\alpha }{{2M }}\le\sin \theta_{\max}\le 1$, $\Xi (\sin{\theta _{\max }})$ can be expressed using (\ref{ps}) and (\ref{pc}) as
\begin{equation}
	\setlength\abovedisplayskip{2pt}
	\setlength\belowdisplayskip{1pt}
	\small
	\begin{aligned}
&\Xi (\sin{\theta _{\max }})\\
&=\alpha \frac{{\left( {1 - \frac{{2{n_{\max }} + 1}}{\eta }} \right)\sin {\theta _{\max }} - \frac{\alpha }{{4M}}\left( {1 - \frac{1}{{{\eta ^2}}}} \right) + \frac{{{n_{\max }}\left( {{n_{\max }} + 1} \right)}}{{{\eta ^2}}}}}{{{{\sin }^2}{\theta _{\max }}M}}\\
	&=\frac{{{\alpha ^2}\left( {\frac{\alpha }{M} - {\eta ^2} - 1} \right)}}{{4{\eta ^2}{M^2}}}{\left( {\frac{1}{{\sin {\theta _{\max }}}}} \right)^2} + \frac{\alpha }{M}\left( {\frac{1}{{\sin {\theta _{\max }}}}} \right) - \frac{\alpha }{M},
	\end{aligned}
	\label{pc1-ps1}
\end{equation}
which is a concave quadratic function of $\frac{1}{\sin\theta_{\max}}$ with $\Xi(\alpha/2M)>0$ and $\Xi(1)<0$. Furthermore, by letting $\Xi (\sin{\theta _{\max }}) = 0$, we obtain $\sin \overline{\theta}  = \frac{{\eta  + \sqrt {{\eta ^2} - \frac{\alpha }{M}({\eta ^2} + 1 - \frac{\alpha }{M})} }}{{2\eta }}$. When $\sin\theta_{\max}\in[\frac{\alpha}{2M}, \overline{\theta}]$, we always have $p(\eta)<p_\mathrm{col}$. When $\sin\theta_{\max}>\overline{\theta}$, $p_\mathrm{col}$ may be slightly smaller than $p(\eta)$. 

Combining all the three cases above, the results in Theorem \ref{appendix b} can be obtained.
	\vspace{-1.5ex}
\bibliographystyle{IEEEtran}
\bibliography{IEEEabrv,ref}
\end{document}